\documentclass[12pt]{article}

\usepackage [latin1]{inputenc}
\usepackage[centertags]{amsmath}
\usepackage{amsmath}
\usepackage{amsfonts}
\usepackage{amssymb}
\usepackage{amsthm}
\usepackage{newlfont}
\usepackage{epsfig}
\usepackage{amscd}
\usepackage{pstricks}
\usepackage{epsfig}
\usepackage{graphicx}
\usepackage{color}
\usepackage{cite}

\newcommand{\beq}{\begin{equation}}
\newcommand{\eeq}{\end{equation}}
\newcommand{\ba}{\begin{array}}
\newcommand{\ea}{\end{array}}
\newcommand{\bea}{\begin{eqnarray}}
\newcommand{\eea}{\end{eqnarray}}

\usepackage[centertags]{amsmath}
\usepackage{amsfonts}
\usepackage{amssymb}
\usepackage{amsthm}
\usepackage{newlfont}
\usepackage{epsfig}
\usepackage{amscd}

\begin{document}

\begin{center}
{\large \sc \bf On the Davey-Stewartson hierarchy: construction by two scalar pseudo-differential operators and compatibility for infinite many flows}

\vskip 20pt

{\large G.Yi $^*$ and X. Liao }

\vskip 20pt

{\it
School of Mathematics, Hefei University of Technology, Hefei 230601, China }

\bigskip

$^*$ Corresponding author:  {\tt ge.yi@hfut.edu.cn}

\bigskip

{\today}

\end{center}

\begin{abstract}
The infinite many symmetries of  Davey-Stewartson (DS) system are closely connected to the integrable deformations of surfaces in a four-dimensional space. In this paper, we give a direct algorithm to construct the expression of the DS hierarchy by two scalar pseudo-differential operators involving partial derivatives.
 \end{abstract}

\section{Introduction}
 The KP (Kadomtsev-Petviashvili) equation \cite{KP1970,AblCla}
 \begin{equation} \label{kp equation}
 (u_{t}+6uu_{x}+u_{xxx})_{x} + 3 \sigma^{2} u_{yy}=0,
 \end{equation}
 and the DS (Davey-Stewartson) system \cite{AblCla,DavSte}
 \begin{subequations} \label{DavSte}
\begin{eqnarray}
\textbf{i} q_{t}+\frac{1}{2}(q_{xx}+\sigma^{2} q_{yy})+\delta q \phi=0,
\end{eqnarray}
\begin{eqnarray}
\sigma^{2}\phi_{yy}-\phi_{xx}+(|q|^{2})_{xx}+\sigma^{2}(|q|^{2})_{yy}=0,
\end{eqnarray}
\end{subequations}
as the most important classical integrable models in (2+1) dimensions, have been extensively studied with many important results obtained \cite{AS1979,Mulase1988,BB1989,Fokas1987,Satsuma1976,CK2010,Kodama2004,DCZ2014,EJK2009,Wazwaz2007,AnkFre,
APPb,Naka1,Naka2,BLMP,ChaWin,Omote,Tajiri,GanLak,FokSan,SanFok,Fokas2009,AKMM2019}. An integrable system is usually associated with a hierarchy of nonlinear partial differential equations defining infinitely many symmetries. This is one of the most important and valuable properties of integrable systems.

The KP hierarchy plays a fundamental role in the theory of integrable systems, a key reason is the clear and explicit definition via Lax equations of pseudo-differential operators \cite{GelDic,SJ1983,Dickey,TK2000,Watanabe,Orlov,AdSh1,AdSh2}.
Let
\begin{equation}
L=\partial+u_{1} \partial^{-1}+u_{2} \partial^{-2}+\cdots,~~~~~~\partial=\frac{\partial}{\partial x},
\end{equation}
be a pseudo-differential operator whose coefficients $u_{i}$ depending on the  spatial coordinate $x$. The KP hierarchy is defined as the following set of equations
\begin{equation}
\frac{\partial L}{\partial t_{n}}=[(L^{n})_{+},L],~~~~~~~~n=1,2,3,\cdots.
\end{equation}
Here the subscript $"+"$ means to take the purely differential part (nonnegative part) of the pseudo-differential operator, while the subscript $"-"$ means to take the negative part. The well-known GD (Gelfand-Dickey) hierarchy is a reduction of the KP hierarchy with the constraint $(L^{m})_{-}=0$ for some natural number $m$.

The KP hierarchy has been generalized to multicomponent cases with scalar pseudo-differential operators replaced by matrix-value ones \cite{Zhang,Satoholland,Dickeymatrix,Leur}. Wu, Zhou, and Lu studied an extension of the KP hierarchy by considering two particular pseudo-differential operators \cite{Wu1,Wu2}. The infinite many symmetries of DS system are closely connected to the 2-component KP hierarchy \cite{konopelmulti,AblCla}. Konopelchenko, Landolfi and Taimanov studied the infinite many symmetries of DS system and pointed out that any symmetry induces an infinite family of geometrically different deformations of tori in $\mathbb{R}^{4}$ preserving the Willmore functional. They defined the DS hierarchy by considering the compatibility of undetermined differential operators in terms of $\partial_{z}$ and $\hat{\partial}_{z}$ \cite{Kono1,Kono2,Taimanov} and gave examples of $t_{2}$ and $t_{3}$ flows. But how to characterize the compatibility for the infinite many equations in the whole hierarchy?

In this paper, we give a direct algorithm to construct the expression of the flows of the DS hierarchy by two scalar pseudo-differential operators involved with $\partial,\hat{\partial}$ and proof the compatibility for these infinite flows. The (1+1) dimensional reduction and some examples are discussed in the final section.

\bigskip
\bigskip
\bigskip

\section{The Davey-Stewartson hierarchy}
Firstly we introduce two scalar pseudo-differential operators
\begin{eqnarray}
L=\partial+\sum_{j=1}^{\infty}u_j\partial^{-j},
\end{eqnarray}
\begin{eqnarray}
\hat{L}=\hat{\partial}+\sum_{j=1}^{\infty}\hat{u}_j\hat{\partial}^{-j},
\end{eqnarray}
 in which the coefficients $u_{j}= u_{j}(t_{mn})$ and $\hat{u_{j}} = \hat{u}_{j}(t_{mn})$ depend on complex variables $t_{mn}$ ($m,n$ are nonnegative integers and $m+n \geq 1$). In particular,  $t_{10} \equiv z,t_{01} \equiv \hat{z}$. Here and hereafter we denote $\partial=\frac{\partial}{\partial z}$, $\hat{\partial}=\frac{\partial}{\partial\hat{z}}$ in this paper.

Similar to the definition of the KP hierarchy, we denote
\begin{subequations} \label{AmBn}
\begin{eqnarray}
A_{m}=(L^{m})_{+},
\end{eqnarray}
\begin{eqnarray}
B_{n}=(\hat{L}^{n})_{+},
\end{eqnarray}
\end{subequations}
and we  introduce
\begin{subequations} \label{tildeAmBn}
\begin{eqnarray}
\tilde{A}_{m}=(\partial^{-1} \circ q \circ A_{m})_{+},
\end{eqnarray}
\begin{eqnarray}
 \tilde{B}_{n}=(\hat{\partial}^{-1}\circ p \circ B_{n})_{+},
\end{eqnarray}
\end{subequations}
here and hereafter the subscript $"+"$ means to take the purely differential part (nonnegative part) of the pseudo-differential operators both in $\partial$ and $\hat{\partial}$, while the subscript $"-"$ means to take the negative part, $q=q(t_{mn})$ and $p=p(t_{mn})$ depend on complex variables $t_{mn}$ are unknown complex-valued functions in this DS hierarchy. The main result of this paper is the following theorem.

\bigskip
\bigskip
\textbf{Theorem.} The compatibility condition of the following linear system
\begin{subequations} \label{DSh}
\begin{eqnarray}
L\varphi=\sigma_{1}\varphi, \label{DSh1}
\end{eqnarray}
\begin{eqnarray}
\hat{L}\psi=\sigma_{2}\psi,
\end{eqnarray}
\begin{eqnarray}
\varphi_{t_{mn}}=A_{m}\varphi+\tilde{B}_{n}\psi,
\end{eqnarray}
\begin{eqnarray}
\psi_{t_{mn}}=\tilde{A}_{m}\varphi+B_{n}\psi,
\end{eqnarray}
\end{subequations}
is equivalent to the equations about the complex-valued functions $q$ and $p$:
\begin{subequations} \label{comequ}
\begin{eqnarray}
q_{t_{mn}}=B_{n}(q)-A^{\ast}_{m}(q), \label{comequ1}
\end{eqnarray}
\begin{eqnarray}
p_{t_{mn}}=A_{m}(p)-B^{\ast}_{n}(p), \label{comequ2}
\end{eqnarray}
\end{subequations}
in which $\sigma_{1}$ and $\sigma_{2}$ are two parameters, $A_{m},B_{n},\tilde{A}_{m},\tilde{B}_{n}$ are defined by (\ref{AmBn})(\ref{tildeAmBn}), $F^{\ast}$ means the adjoint operator to $F$.
This compatible system (\ref{DSh}) is defined as DS (Davey-Stewartson) hierarchy, and the infinite number of (2+1) dimensional nonlinear partial differential equations (\ref{comequ}) are corresponding flow equations of the DS hierarchy.

\bigskip

\textbf{Remark 1.} Without the effect of $\tilde{A}_{m}, \tilde{B}_{n}$, the above system is nothing but two separated KP hierarchies. In fact, $\tilde{A}_{m}, \tilde{B}_{n}$ connect the two KP hierarchies and keep the compatibility. This is the key point for constructing this DS hierarchy.

\bigskip

\textbf{Remark 2.} Notice that $\varphi_{t_{10}}=\varphi_{z},\psi_{t_{01}}=\psi_{\hat{z}}$. This implies $\frac{\partial}{\partial t_{10}}=\partial,\frac{\partial}{\partial t_{01}}=\hat{z}$.
This is the reason we identify $t_{10}$ with $z$ and $t_{01}$ with $\hat{z}$ in this paper. Therefore, one obtains the following Dirac system
\begin{subequations} \label{Dirac}
\begin{eqnarray}
\varphi_{\hat{z}}=\varphi_{t_{01}}=\tilde{B}_{1}\psi=p\psi,
\end{eqnarray}
\begin{eqnarray}
\psi_{z}=\psi_{t_{10}}=\tilde{A}_{1}\varphi=q\varphi.
\end{eqnarray}
\end{subequations}
Then the compatibility condition of linear system (\ref{DSh}) reads as follows
\begin{subequations} \label{DirDS}
\begin{eqnarray}  \label{DirDS1}
\varphi_{\hat{z}}=p\psi,
\end{eqnarray}
\begin{eqnarray} \label{DirDS2}
\psi_{z}=q\varphi,
\end{eqnarray}
\begin{eqnarray}
\varphi_{t_{mn}}=A_{m}\varphi+\tilde{B}_{n}\psi, \label{DirDS3}
\end{eqnarray}
\begin{eqnarray} \label{DirDS4}
\psi_{t_{mn}}=\tilde{A}_{m}\varphi+B_{n}\psi.
\end{eqnarray}	
\end{subequations}
The above hierarchy is called the  DS hierarchy for the reason that its $t_{22}$ flow is the well-known DS system (\ref{DavSte}).

\bigskip
\bigskip
To prove this main theorem, we need the following two lemmas.

\textbf{Lemma 1.}
The pseudo-differential operators $L$ and $\hat{L}$ satisfy the following structure equations respectively
\begin{subequations} \label{Laxe}
\begin{eqnarray}
\frac{\partial L^{m}}{\partial\hat{z}}+[L^{m},R]=0, \label{Laxe1}
\end{eqnarray}
\begin{eqnarray}
\frac{\partial\hat{L}^{n}}{\partial z}+[\hat{L}^{n},\hat{R}]=0 ,\label{Laxe2}
\end{eqnarray}	
\end{subequations}
in which
\begin{equation}
R=p \circ \partial^{-1} \circ q,\quad \hat{R}=q \circ \hat{\partial}^{-1} \circ p.
\end{equation}
Correspondingly, $A_{m}$ and $B_{n}$ satisfy
\begin{subequations}
\begin{eqnarray} \label{struequ1}
\frac{\partial A_{m}}{\partial\hat{z}}+[A_{m},R]_{+}=0,
\end{eqnarray}
\begin{eqnarray} \label{struequ2}
\frac{\partial B_{n}}{\partial z}+[B_{n},\hat{R}]_{+}=0.
\end{eqnarray}
\end{subequations}
\textbf{Proof.}
In fact, the equation (\ref{DSh1}) yields $L^{m}\varphi=\sigma_{1}^{m}\varphi$.
By taking the derivative of this equation with respect to $\hat{z}$, one obtains
\begin{equation*}
\frac{\partial L^{m}}{\partial\hat{z}}\varphi+L^{m}\varphi_{\hat{z}}-\sigma_{1}^{m}\varphi_{\hat{z}}=0,
\end{equation*}
which leads to
\begin{equation*}
\left(\frac{\partial L^{m}}{\partial\hat{z}}+L^{m}\circ p \circ \partial^{-1}\circ q-p\circ \partial^{-1} \circ q \circ L^{m}\right)\varphi=0.
\end{equation*}
Hence,(\ref{Laxe1}) is true. The discussion for (\ref{Laxe2}) is similar. By taking the differential part (nonnegative part) of  (\ref{Laxe1}) and (\ref{Laxe2}), one obtains (\ref{struequ1}) and (\ref{struequ2}).

~~~~~~~~~~~~~~~~~~~~~~~~~~~~~~~~~~~~~~~~~~~~~~~~~~~~~~~~~~~~~~~~~~~~~~~~~~~~~~~~~~~~~~~~~~~$\square$

\bigskip

\textbf{Remark 3.} From the structure equations (\ref{Laxe}), we can deduce that all the coefficients $u_{j}$ and $\hat{u}_{j}$ of the pseudo-differential operators  $L$ and $\hat{L}$ depend on the two complex-valued functions $p,q$ and their derivatives or integrals with respect to the independent variables $z,\hat{z}$. The leading terms of $L$ and $\hat{L}$ can be obtained directly by (\ref{Laxe}) as follows
\begin{subequations}
\begin{eqnarray}\nonumber
\quad u_{1}=-\partial_{\hat{z}}^{-1}\left((pq)_{z}\right),\qquad \quad u_{2}=-\partial_{\hat{z}}^{-1}\left((q_{z}p)_{z}\right),\qquad\\ u_{3}=-\partial_{\hat{z}}^{-1}\left((pq_{zz})_{z}\right)-\partial_{\hat{z}}^{-1}\left((pq)_{z}\partial_{\hat{z}}^{-1}((pq)_{z})\right)
+\partial_{\hat{z}}^{-1}\left(pq\partial_{\hat{z}}^{-1}\left((pq)_{zz}\right)\right),\cdots,
\end{eqnarray}
\begin{eqnarray}\nonumber
\quad \hat{u}_{1}=-\partial_{z}^{-1}\left((pq)_{\hat{z}}\right),\qquad \quad \hat{u}_{2}=-\partial_{z}^{-1}\left((p_{\hat{z}}q)_{\hat{z}}\right),\qquad\\
\hat{u}_{3}=-\partial_{z}^{-1}\left((p_{\hat{z}\hat{z}}q)_{\hat{z}}\right)-\partial_{z}^{-1}\left((pq)_{\hat{z}}\partial_{z}^{-1}((pq)_{\hat{z}})\right)
+\partial_{z}^{-1}\left(pq\partial_{z}^{-1}\left((pq)_{\hat{z}\hat{z}}\right)\right),\cdots.
\end{eqnarray}
\end{subequations}

\bigskip

\textbf{Lemma 2.}
The two pairs of differential operators $A_{m}$, $\tilde{A}_{m}$  and $B_{n}$,$\tilde{B}_{n}$ satisfy
\begin{subequations}
\begin{eqnarray} \label{relationAm}
\partial \circ \tilde{A}_{m}-q \circ A_{m}=-A^{\ast}_{m}(q),
\end{eqnarray}
\begin{eqnarray} \label{relationBn}
\hat{\partial} \circ \tilde{B}_{n}-p \circ B_{n}=-B^{\ast}_{n}(p).
\end{eqnarray}
\end{subequations}

\textbf{Proof.} In fact, $A_{m}$ defined in (\ref{AmBn}) reads as
\begin{eqnarray}\nonumber
A_{m}=\sum_{k=0}^{m}a_{k}\partial^{k},
\end{eqnarray}
where $a_{m}=1,a_{m-1}=0$ and $a_{k} (k=0,1\cdots,m-2)$ are differential polynomials in $u_{j} (j=1,2,\cdots,m-1)$. Therefore,
\begin{eqnarray}\nonumber
\partial \circ \tilde{A}_{m}-q \circ A_{m}&=&\partial \circ (\partial^{-1} \circ q \circ A_{m})_{+}-\partial \circ \partial^{-1} \circ q \circ A_{m}\\\nonumber
&=&-\partial \circ (\partial^{-1} \circ q \circ A_{m})_{-}\\\nonumber
&=&-\partial \circ \left(\partial^{-1}\circ\left(\sum_{k=0}^{m}(-1)^{k}(qa_{k})^{(k)}\right)\right)\\\nonumber
&=&-A^{\ast}_{m}(q),
\end{eqnarray}
where $(qa_{k})^{(k)}=\frac{\partial ^{k} (qa_{k})}{\partial z^{k}}$. The proof of (\ref{relationBn}) is similar.

~~~~~~~~~~~~~~~~~~~~~~~~~~~~~~~~~~~~~~~~~~~~~~~~~~~~~~~~~~~~~~~~~~~~~~~~~~~~~~~~~~~~~~~~~~~$\square$

\bigskip

With the help of the two lemmas, we come to the proof of the main theorem.

\textbf{Proof of the theorem.}
In fact, the compatibility condition of (\ref{DirDS1}) and (\ref{DirDS3}) reads as
\begin{eqnarray} \nonumber
\frac{\partial^{2}\psi}{\partial z\partial t_{mn}}=\frac{\partial^{2}\psi}{\partial t_{mn}\partial z}.
\end{eqnarray}
By direct calculation, one obtains
\begin{eqnarray}\nonumber
\frac{\partial^{2}\psi}{\partial z\partial t_{mn}}&=&q_{t_{mn}}\varphi+qA_{m}\varphi+q\tilde{B}_{n}\psi,\nonumber
\end{eqnarray}
\begin{eqnarray}\nonumber
\frac{\partial^{2}\psi}{\partial t_{mn}\partial z}=\partial(\tilde{A}_{m}\varphi)+\partial(B_{n}\psi).
\end{eqnarray}
Therefore,
\begin{eqnarray} \label{compa1}
\frac{\partial^{2}\psi}{\partial z\partial t_{mn}}-\frac{\partial^{2}\psi}{\partial t_{mn}\partial z}&=&q_{t_{mn}}\varphi+\left( qA_{m}\varphi-\partial(\tilde{A}_{m}\varphi)\right)+\left(q\tilde{B}_{n}\psi-\partial(B_{n}\psi)\right)=0.\nonumber\\
&&
\end{eqnarray}
By the virtue of (\ref{relationAm}) in Lemma 2, one obtains
\begin{eqnarray}\nonumber \label{part1}
qA_{m}\varphi-\partial(\tilde{A}_{m}\varphi)&=&\left(q\circ A_{m}-\partial \circ \tilde{A}_{m}\right)(\varphi)\\
&=&A^{\ast}_{m}(q)\varphi.
\end{eqnarray}
The other part in (\ref{compa1}) can be simplified by direct calculation as follows
\begin{eqnarray}\nonumber \label{part2}
q\tilde{B}_{n}\psi-\partial(B_{n}\psi)&=&q\tilde{B}_{n}\psi-\frac{\partial B_{n}}{\partial z}\psi-B_{n}(q\varphi)\\\nonumber
&=&q\tilde{B}_{n}\psi-\frac{\partial B_{n}}{\partial z}\psi-B_{n}\circ q\circ \hat{\partial}^{-1}\circ p(\psi)\\\nonumber
&=&\left( q\circ\tilde{B}_{n}-(B_{n}\circ q\circ\hat{\partial}^{-1}\circ p)_{+}-\frac{\partial B_{n}}{\partial z} \right)(\psi)-\left(B_{n}\circ q\circ\hat{\partial}^{-1}\circ p \right)_{-}(\psi)\\\nonumber
&=&\left(\frac{\partial B_{n}}{\partial z}+[B_{n},\hat{R}]_{+}\right)(\psi)-\left( B_{n}(q)\circ\hat{\partial}^{-1}\circ p \right)(\psi)\\
&=&\left( \frac{\partial B_{n}}{\partial z}+[B_{n},\hat{R}]_{+} \right)(\psi)-B_{n}(q)\varphi.
\end{eqnarray}
Then the structure equation (\ref{struequ2}) in Lemma 1 gives
\begin{eqnarray} \label{part21}
q\tilde{B}_{n}\psi-\partial(B_{n}\psi)=-B_{n}(q)\varphi.
\end{eqnarray}
Therefore the compatibility equation (\ref{compa1}) can be simplified as
\begin{eqnarray}
\frac{\partial^{2}\psi}{\partial z\partial t_{mn}}-\frac{\partial^{2}\psi}{\partial t_{mn}\partial z}=\left( q_{t_{mn}}+A^{\ast}_{m}(q)-B_{n}(q) \right)\varphi=0,
\end{eqnarray}
which leads to the flow equation (\ref{comequ1}).

Similarly, by considering the compatibility condition
\begin{eqnarray} \nonumber
\frac{\partial^{2}\varphi}{\partial\hat{z}\partial t_{mn}}=\frac{\partial^{2}\varphi}{\partial
t_{mn}\partial\hat{z}},
\end{eqnarray}
one obtains the flow equation (\ref{comequ2}).

~~~~~~~~~~~~~~~~~~~~~~~~~~~~~~~~~~~~~~~~~~~~~~~~~~~~~~~~~~~~~~~~~~~~~~~~~~~~~~~~~~~~~~~~~~~$\square$
\bigskip
\bigskip

\section{Examples and (1+1) dimensional reduction}
Some examples from the DS hierarchy and the (1+1) dimensional reduction are presented below.

\bigskip
\textbf{Example 1:}

By taking $m=n=2$, then
\begin{subequations}
\begin{eqnarray}
A_{2}=\partial^{2}+2u_{1}=\partial^{2}-2\partial_{\hat{z}}^{-1}\left((pq)_{z}\right),
\end{eqnarray}
\begin{eqnarray}
B_{2}=\hat{\partial}^{2}+2\hat{u}_{1}=\hat{\partial}^{2}-2\partial_{z}^{-1}\left((pq)_{\hat{z}}\right),
\end{eqnarray}
\begin{eqnarray}
\tilde{A_{2}}=q\partial-q_{z},
\end{eqnarray}
\begin{eqnarray}
\tilde{B_{2}}=p\hat{\partial}-p_{\hat{z}}.
\end{eqnarray}
\end{subequations}
Therefore, linear system (\ref{DirDS}) reads as
\begin{subequations} \label{DSlax}
\begin{eqnarray}
\varphi_{\hat{z}}=p\psi,
\end{eqnarray}
\begin{eqnarray}
\psi_{z}=q\varphi,
\end{eqnarray}
\begin{eqnarray}
\varphi_{t_{22}}=\left(\partial^{2}-2\partial_{\hat{z}}^{-1}((pq)_{z})\right)\varphi+\left(p\hat{\partial}-p_{\hat{z}}\right)\psi,
\end{eqnarray}
\begin{eqnarray}
\psi_{t_{22}}=\left(q\partial-q_{z}\right)\varphi+\left(\hat{\partial}^{2}-2\partial_{z}^{-1}((pq)_{\hat{z}})\right)\psi.
\end{eqnarray}		
\end{subequations}
Then the following system
\begin{subequations} \label{Dsequ}
\begin{eqnarray}
q_{t_{22}}=B_{2}(q)-A^{\ast}_{2}(q) = q_{\hat{z}\hat{z}}-q_{zz}-\phi q,
\end{eqnarray}
\begin{eqnarray}
p_{t_{22}}=A_{2}(p)-B^{\ast}_{2}(p) = p_{zz}-p_{\hat{z}\hat{z}}+\phi p,
\end{eqnarray}
\begin{eqnarray}
\phi_{z\hat{z}}=2(pq)_{\hat{z}\hat{z}}-2(pq)_{zz},
\end{eqnarray}
\end{subequations}
in the form (\ref{comequ}) arising from (\ref{DSlax}) is nothing but the well-known DS system with $p=\bar{q}$.

\bigskip

\textbf{Example 2:}

By taking $m=n=3$,then
\begin{subequations}
\begin{eqnarray}\nonumber
A_{3}&=&\partial^{3}+3u_{1}\partial+3u_{2}+3u_{1z}\\
&=&\partial^{3}-3\partial_{\hat{z}}^{-1}\left((pq)_{z}\right)\partial-3\partial_{\hat{z}}^{-1}\left((q_{z}p)_{z}\right)
-3\partial_{\hat{z}}^{-1}\left((pq)_{zz}\right),
\end{eqnarray}
\begin{eqnarray}\nonumber
B_{3}&=&\hat{\partial}^{3}+3\hat{u}_{1}\hat{\partial}+3\hat{u}_{2}+3\hat{u}_{1\hat{z}}\\
&=&\hat{\partial}^{3}-3\partial_{z}^{-1}\left((pq)_{\hat{z}}\right)\hat{\partial}-3\partial_{z}^{-1}\left((p_{\hat{z}}q)_{\hat{z}}\right)
-3\partial_{z}^{-1}\left((pq)_{\hat{z}\hat{z}}\right),
\end{eqnarray}
\begin{eqnarray}
\tilde{A}_{3}&=&q \partial^{2}-q_{z} \partial+q_{zz}+3qu_{1},
\end{eqnarray}
\begin{eqnarray}
\tilde{B}_{3}&=&p  \hat{\partial}^{2}-p_{\hat{z}} \hat{\partial}+p_{\hat{z}\hat{z}}+3p \hat{u}_{1}.
\end{eqnarray}
\end{subequations}

Correspondingly, the adjoint operators read as
\begin{subequations}
\begin{eqnarray}\nonumber
A^{\ast}_{3}&=&-\partial^{3}-3u_{1}\partial+3u_{2}+3u_{1z}\\
&=&-\partial^{3}+3\partial_{\hat{z}}^{-1}\left((pq)_{z}\right)\partial-3\partial_{\hat{z}}^{-1}\left((q_{z}p)_{z}\right)
-3\partial_{\hat{z}}^{-1}\left((pq)_{zz}\right),
\end{eqnarray}
\begin{eqnarray}\nonumber
B^{\ast}_{3}&=&-\hat{\partial}^{3}-3\hat{u}_{1}\hat{\partial}+3\hat{u}_{2}+3\hat{u}_{1\hat{z}}\\
&=&-\hat{\partial}^{3}+3\partial_{z}^{-1}\left((pq)_{\hat{z}}\right)\hat{\partial}-3\partial_{z}^{-1} \left((p_{\hat{z}}q)_{\hat{z}}\right)
-3\partial_{z}^{-1}\left((pq)_{\hat{z}\hat{z}}\right).
\end{eqnarray}
\end{subequations}
Then the communication of linear system (\ref{DirDS}) is equivalent to the following integrable system
\begin{subequations}
\begin{eqnarray}
q_{t_{33}}=B_{3}(q)-A^{\ast}_{3}(q),
\end{eqnarray}
\begin{eqnarray}
p_{t_{33}}=A_{3}(p)-B^{\ast}_{3}(p).
\end{eqnarray}
\end{subequations}

\bigskip
\bigskip

\textbf{Example 3:}

By taking $m=2,n=3$, then the communication condition of the linear system (\ref{DirDS}) reads as
\begin{subequations}
\begin{eqnarray}
q_{t_{23}}=B_{3}(q)-A^{\ast}_{2}(q),
\end{eqnarray}
\begin{eqnarray}
p_{t_{23}}=A_{2}(p)-B^{\ast}_{3}(p).
\end{eqnarray}
\end{subequations}

\bigskip
\bigskip

Now, we consider the (1+1) dimensional reduction of the DS hierarchy which includes the important integrable model NLS (nonlinear Schr$\ddot{o}$dinger) equation.  By considering the conjugate independent variables $z=x+\textbf{i}y,\hat{z}=x-\textbf{i}y$ and the reduced condition $\partial=\textbf{i}\hat{\partial}$, i.e., $x=-y$, one obtains the following (1+1) reduction of the linear system (\ref{DirDS}) under the transform $p\to\frac{1-\textbf{i}}{2} p,q\to\frac{1+\textbf{i}}{2} q$
\begin{subequations} \label{reduced DS}
\begin{eqnarray}
\varphi_{x}=p\psi,
\end{eqnarray}
\begin{eqnarray}
\psi_{x}=q\varphi,
\end{eqnarray}
\begin{eqnarray}
\varphi_{t_{mn}}=\alpha_{m}\varphi+\tilde{\beta}_{n}\psi,
\end{eqnarray}
\begin{eqnarray}
\psi_{t_{mn}}=\tilde{\alpha}_{m}\varphi+\beta_{n}\psi,
\end{eqnarray}		
\end{subequations}
where $\alpha_{m}$, $\beta_{n}$ , $\tilde{\alpha}_{m}$ and $\tilde{\beta}_{n}$  are differential operators in terms of $\partial_{x}$. The following nontrivial example of this reduced hierarchy is the NLS equation.

\bigskip
\bigskip
\bigskip
\bigskip

\textbf{Example 4:}

By taking $m=n=2$, the linear system (\ref{reduced DS}) reads as
\begin{subequations} \label{reduced ex1}
\begin{eqnarray}
\varphi_{x}=p\psi,
\end{eqnarray}
\begin{eqnarray}
\psi_{x}=q\varphi,
\end{eqnarray}
\begin{eqnarray}
\varphi_{t_{22}}=\left(\frac{\textbf{i}}{2}\partial_{x}^{2}-\textbf{i}pq\right)\varphi
+\left(-\frac{\textbf{i}}{2}p\partial_{x}+\frac{\textbf{i}}{2}p_{x}\right)\psi,
\end{eqnarray}
\begin{eqnarray}
\psi_{t_{22}}=\left(\frac{\textbf{i}}{2}q\partial_{x}-\frac{\textbf{i}}{2}q_{x}\right)\varphi
+\left(-\frac{\textbf{i}}{2}\partial_{x}^{2}+\textbf{i}pq\right)\psi.
\end{eqnarray}		
\end{subequations}
The compatibility condition of (\ref{reduced ex1}) is equivalent to
\begin{subequations}
\begin{eqnarray}
\textbf{i}q_{t}-q_{xx}+2pq^{2}=0,
\end{eqnarray}
\begin{eqnarray}
\textbf{i}p_{t}+p_{xx}-2pq^{2}=0,
\end{eqnarray}
\end{subequations}
which is exactly the classical NLS equation with $p=\pm \bar{q}$.

\bigskip

\section{Outlook}
Research on the dispersionless integrable systems (integrable systems of hydrodynamic type) which arise from the commutation condition of vector fields Lax pairs, is an important subject. One important kind of dispersionless integrable systems comes from the semiclassical limit (dispersionless limit) of the classical integrable systems. In \cite{yi1}, we discussed the semiclassical limit of the DS system (\ref{DavSte}) and the relevant nonlinear Riemann-Hilbert problem. In \cite{yi2}, we defined a new class of dispersionless integrable systems called dDS (dispersionless Davey-Stewartson) hierarchy. In fact, the semiclassical limit (dispersionless limit) of the DS hierarchy (\ref{DSh})(\ref{DirDS}) is closely connected to the dDS hierarchy. We will show the details in the next separated paper.

\bigskip
\bigskip
\bigskip

\textbf{Acknowledgements:} This work has been supported by the Key Laboratory Foundation (No.6142209180306) and National Natural Science Foundation of China (No. 11501222).

\end{document}